\documentclass[iop,twocolumn,apj,numberedappendix,appendixfloats]{emulateapj}
\usepackage{apjfonts,color}
\usepackage[pagebackref=false,letterpaper=true,colorlinks=true,citecolor=blue,linkcolor=blue,breaklinks=true,bookmarks=true]{hyperref}

\usepackage{mathrsfs}
\usepackage{epsfig}
\usepackage{graphicx}
\usepackage{mathtools}
\usepackage{color}
\usepackage{url}
\usepackage{hyperref}
\usepackage{footnote}

\usepackage{natbib}
\usepackage{multirow}
\usepackage{cleveref}

\newcommand{\jcap}{J. Cosmol. Astropart. Phys.}

\newcommand{\bx}{{\bf x}}
\newcommand{\bk}{{\bf k}}

\begin{document}
\title{Polarization of the Cosmic Infrared Background Fluctuations}

\author{Chang Feng}
\email{changf@illinois.edu}
\affiliation{Department of Physics, University of Illinois at Urbana-Champaign, 1110 West Green Street, Urbana, Illinois, 61801, USA}

\author{Gilbert Holder}
\affiliation{Department of Physics, University of Illinois at Urbana-Champaign, 1110 West Green Street, Urbana, Illinois, 61801, USA}
\affiliation{Department of Astronomy, University of Illinois at Urbana-Champaign, 1002 West Green Street, Urbana, Illinois, 61801, USA}
\affiliation{Canadian Institute for Advanced Research, Toronto, Ontario M5G 1M1, Canada}

\begin{abstract}
The cosmic infrared background (CIB) is slightly polarized. Polarization directions of individual galaxies could be aligned with tidal fields around galaxies, resulting in nonzero CIB polarization. We use a linear intrinsic alignment model to theoretically predict angular correlations of the CIB polarization fluctuations and find that electriclike and curl-like ($B$-mode) polarization modes are equally generated with power four orders of magnitude less than its intensity. The CIB $B$-mode signal is negligible and not a concerning foreground for the inflationary $B$-mode searches at nominal frequencies for cosmic microwave background measurements, but could be detected at submillimetre wavelengths by future space missions.
\end{abstract}

\maketitle

\section{Introduction}
\label{intro}

The angular correlation pattern of cosmic microwave background (CMB) polarization fluctuations was predicted shortly after the CMB discovery in 1965 \citep{1968ApJ...153L...1R,1979ApJ...232..341N, 1980PhRvL..44.1433N, 1984ApJ...285L..45B, 1985ApJ...290....1T, 1985SvA....29..607P}. In 2001, the Degree Angular Scale Interferometer (DASI) experiment for the first time detected an electriclike component ($E$-mode) of the predicted polarization signal \citep{2002Natur.420..772K} and opened up a window for the CMB polarization science. Since then, ground-based CMB experiments including the Background Imaging of Cosmic Extragalactic Polarization (BICEP) \citep{2010ApJ...711.1123C}, the POLARization of the Background Radiation (POLARBEAR) \citep{2014ApJ...794..171P}, Atacama Cosmology Telescope (ACT) \citep{2014JCAP...10..007N} and South Pole Telescope (SPT) \citep{2015ApJ...805...36C} have all detected the $E$-mode polarization with great precision. From space, the Planck experiment has made the most precise measurement of the $E$ mode to date \citep{2016A&A...594A..11P}. As also predicted by perturbation theory, there is the existence of another, curl-like, component of the polarization signal --- the so-called $B$ mode, which can be created by both inflationary gravitational waves and gravitational lensing of the $E$-mode component, corresponding to $B$-mode fluctuations at large and small angular scales, respectively. The lensing $B$-mode, caused by large-scale structure, has been detected by many experiments \citep{2013PhRvL.111n1301H, 2014ApJ...794..171P, 2014PhRvL.113b1301A,2014PhRvL.112x1101B,2015ApJ...807..151K,2017ApJ...848..121P}. With sophisticated low-temperature detection technologies, cosmologists have gained a good understanding of the CMB on linear scales, although the inflationary $B$ modes remain a mystery.

Future high sensitivity and high resolution (HSHR) CMB experiments \citep{2019arXiv190704473A, 2019arXiv190610134S, 2018SPIE10708E..04G, 2014SPIE.9153E..1PB} will enter a new era when the CMB will be measured at very small and even nonlinear scales with unprecedented precision, and the secondary CMB fluctuations --- radio galaxies, the thermal/kinetic Sunyaev-Zel'dovich (tSZ/kSZ) effects and the cosmic infrared background (CIB) --- will be precisely measured. As one of the precursors to the HSHR experiments, the SPT collaboration analyzed 2500 ${\rm deg}^2$ data taken at 90, 150 and 220 GHz and found that the CIB emission due to the dusty star-forming galaxies is much stronger than the tSZ and kSZ at scales smaller than $4'$ \citep{2015ApJ...799..177G}. Above 95 GHz the radio galaxies have little contributions to the secondaries \citep{2014A&A...566A...5I}. Moreover, the CIB intensity anisotropies at high CMB frequencies --- 217, 353, 545 and 857 GHz --- are also measured by the Planck satellite and are found to dominate the primary fluctuations \citep{2014A&A...571A..30P}. 

The secondary fluctuations can also be polarized. Fluctuations from polarized synchrotron radiation from radio galaxies are negligible at $>150$ GHz \citep{2014A&A...566A...5I} and the power of polarized tSZ signal is theoretically predicted to be $\mathcal{O}(10^{-5}) \mu {\rm K}^2$ and behave like white noise \citep{2018JCAP...04..034D}. The CIB dominates the secondaries in intensity at small scales but its polarization property has not been investigated. 
 
CIB polarization arises from polarized thermal dust emissions of individual galaxies. The dust emission of individual galaxies is known to be polarized and different physical mechanisms have been proposed to explain it \citep{1966ApJ...144..318S,1988ApL&C..26..263H, 2018PhRvL.121b1104K}. One basic theory is that asymmetric dust grains are aligned with the magnetic field of the host galaxy, leading to polarized emission that is perpendicular to the alignment direction \citep{1966ApJ...144..318S,1988ApL&C..26..263H}. Turbulence is also thought to introduce Galactic dust polarization, as found from numerical simulations \citep{2018PhRvL.121b1104K}. Although differences lie in various theories, the average polarization fraction of individual galaxy is theoretically estimated to be $1-2\%$ of its intensity \citep{1966ApJ...144..318S, 1988QJRAS..29..327H}. 
From the Planck 353 GHz data, the polarization of the Galactic dust indicates that the dust is polarized at $\sim$10\% level \citep{2015A&A...576A.104P, 2018arXiv180104945P}. 

The CIB arises from the superposition of emissions from many galaxies. The CIB polarization should be less than individual galaxy polarization as misalignments between galaxies will lead to the averaging down of polarization fluctuations. Thus the CIB is usually assumed to be unpolarized and fully described by its intensity fluctuations. However, it is unclear how much polarization remains after this averaging. The polarization vectors of galaxies trace magnetic fields and directions of magnetic fields are correlated with galaxy morphologies, such as galaxy shapes \citep{1958ApJ...128....9H, 1994A&A...284..777G, 2003A&A...398..937B, 2008ApJ...677L..17C, 2011MNRAS.412.2396F}. Galaxy shapes are known to be correlated with local tidal fields \citep{2009ApJ...694L..83O, 2009ApJ...694..214O, 2015MNRAS.450.2195S, 2018MNRAS.478..711M, 2019A&A...624A..30J}. Therefore, galaxy polarization vectors can be expected to be preferentially aligned with the tidal fields generated by large scale structure. 

If the CIB polarization signal is detectable, it would complement the current polarization surveys with much shorter wavelengths and will open up a new window on structure formation.
If the CIB polarization is not curl-free but contains a $B$-mode signal, it could become a new challenge to CMB inflationary $B$-mode searches, depending on its strength. As a science return, it is even possible that the CIB polarization will bring new cosmological information. In this work, we will theoretically calculate the CIB polarization signal with the linear alignment model proposed in \cite{2004PhRvD..70f3526H}.  \\

\section{Intrinsic alignment induced polarization}
\label{iamodel}
The CIB polarization requires a long-range coherence mechanism between galaxies, which could be caused by galaxy intrinsic alignment (IA). A spatial coupling $T(\bx)\gamma(\bx)$ between the CIB intensity $T(\bx)$ and the tidal field $\gamma(\bx)$ generates the CIB polarization. Here $\bx$ denotes space coordinates.  Theoretical models for the linear intrinsic alignment (IA) are proposed in \cite{2004PhRvD..70f3526H}. The tidal fields are related to the gravitational potential $\Psi(\bx)$
\begin{equation}
{\boldsymbol \gamma}({\bf x})=-\frac{C_1}{4\pi G}(\nabla^2_x-\nabla^2_y, 2\nabla_x\nabla_y)\Psi({\bf x}),
\end{equation}

and in the Fourier domain, it is 
\begin{equation}
\gamma(\bk)=-\frac{C_1}{4\pi G}e^{i2\phi_{\bk_{\perp}}}\Psi({\bk}).
\end{equation}
Here $\bk_{\perp}=(k_x, k_y)$, $\bk_{\parallel}=k_z$, $C_1$ is a free parameter and $\nabla$ is a covariant derivative. Rotational invariance is imposed on the tidal field by a phase factor $e^{i2\phi_{\bk_{\perp}}}$ so the tidal field can be decomposed into the Stokes parameters via $\gamma=\gamma^{Q}+i\gamma^{U}$. Specifically, $\phi_{\bk_{\perp}}$ is a projected angle between $k_x$ and $k_y$ on the $\bk_{\perp}$ plane so the phase factor $e^{i2\phi_{\bk_{\perp}}}=(k_x^2-k_y^2)/k_{\perp}^2+i2k_xk_y/k_{\perp}^2$ where $|\bk_{\perp}|=k_{\perp}$. The gravitational potential is created by the matter distribution $\delta(\bx)$ via the Poisson equation $\nabla^2\Psi({\bf x})=-4\pi G\bar \rho/\bar Da^2\delta({\bf x})$, where $a$ is the cosmological scale factor, $\bar \rho$ is the matter density and $D$ is a growth factor. In this work, we only consider the linear alignment model, and assume that the higher order corrections, such as the quadratic alignment, are negligible.

The CIB is the emission from dust surrounding star-forming regions in distant galaxies. Conceivably, the total polarization signal is $T({\bf x})\langle p\rangle$, where $\langle p\rangle$ is an averaged polarization fraction for an individual source. We assume a polarization fraction $\langle p\rangle=1\%$. For a galaxy, the polarization $Q+iU$ is described by the Stokes parameters $Q$ and $U$, and the angle of the direction is $\alpha=1/2{\arctan}(\gamma^U/\gamma^Q)$. If the polarization direction is randomly aligned, the polarization signal from the extragalactic dust will behave like white noise. However, if the polarization vector at each location is aligned with the local tidal field which simultaneously modulates its polarization intensity, the polarization signal of the extragalactic dust will behave differently from white noise. 

We begin with the IA-induced polarization model    
\begin{equation}
T^P(\bx)=T({\bf x})\langle p\rangle\gamma({\bf x}).\label{iamodel}
\end{equation}
The CIB temperature modulation by the IA field is the key to generating a polarization signal. Next, we will use Eq. (\ref{iamodel}) to predict the CIB polarization power spectra. We assume that the CIB emissivity is linearly proportional to the underlying density field, i.e., $T(\bx)=b_g\bar T_0\delta(\bx)$. The galaxy bias $b_g$ is fixed by the Planck CIB temperature power spectrum at 857 GHz, and the mean CIB intensity $\bar T_0$ is calculated by integrating the flux distribution up to a certain flux threshold.

Expanded by plane waves, Eq. (\ref{iamodel}) becomes
\begin{equation}
T^P(\bk)=F(z)\langle p\rangle\int \frac{d^3k_1}{(2\pi)^3}e^{i2\phi_{{\bf k_{2,\perp}}}}\delta(\bk_1)\delta(\bk_2),
\end{equation}
where $\bk=\bk_1 + \bk_2$ and $e^{i2\phi_{\bk_{\perp}}}=f_A(\bk)+if_B(\bk)$. We adopt a prefactor $F(z)=-AC_1\rho_{\rm crit}\Omega_m/D(z)$ following the definition of \cite{2011A&A...527A..26J,2013MNRAS.432.2433H}, which is slightly different from the original definition in \cite{2004PhRvD..70f3526H}. Here $A$ and $C_1$ are both free parameters, $\rho_{\rm crit}$ is the critical density today and $D$ is the growth factor normalized to unity today. It is seen from this equation that the CIB polarization essentially arises from a two-point matter density correlation. The HSHR experiments measure a projected CIB field which is an integration of all redshift slices weighted by a redshift distribution $W_{\nu}(z)$, i.e.,
\begin{equation}
T^P({\bf n})=\int d\chi W_{\nu}^{\rm CIB}(z)T^P(\chi{\bf n}),
\end{equation}
and the CIB temperature and polarization angular power spectra at frequency $\nu$ are derived using the Limber approximation $k=\ell/\chi$, i.e.,
\begin{equation}
C_{\ell}^{XX, (\nu\nu)}=\int\frac{d\chi}{\chi^2}W^{\rm CIB}_{\nu}(z)W^{\rm CIB}_{\nu}(z)P^{XX}(k,z).
\end{equation}
Here $X=\{T, E, B\}$, $\bf n$ is a direction in the sky, $\chi$ is the comoving distance, and the CIB redshift distribution $W^{\rm CIB}_{\nu}(z)$ is taken from \cite{2013ApJ...772...77V}. 3D power spectra are $P^{TT}=b_g^2P^{m}$ for temperature and $P^{EE/BB}=\langle T^P(\bk)T^{P \ast}(\bk)\rangle$ for polarization \citep{2004PhRvD..70f3526H}, i.e., 

\begin{figure*}
\includegraphics[trim=0 0 0 0, width=19cm, height=9.5cm]{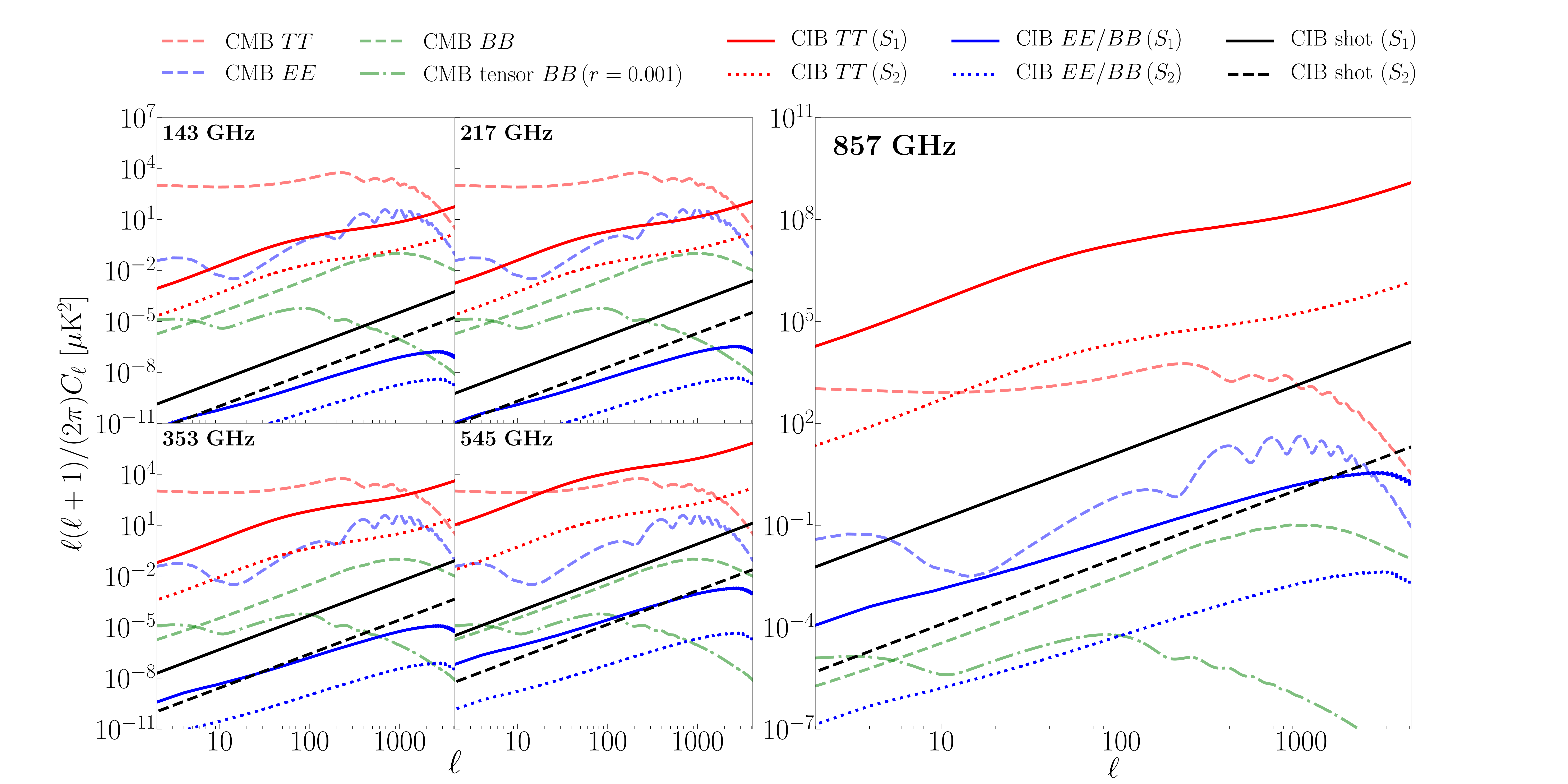}
\caption{(Left) CIB polarization power spectra with respect to the observing frequencies. The CIB polarization is nonzero and its power ($\sqrt{\ell(\ell+1)/(2\pi) C_{\ell}}$) is about four orders of magnitude lower than its intensity. Also the polarization power is equally divided into $E$ and $B$ modes, and is not distributed like white noise. The polarization power increases as the intensity increases. The CIB $B$-mode signal is much fainter than the CMB inflationary $B$-mode at frequencies $\nu<353$ GHz. The CMB tensor $B$-mode is generated with a tensor-to-scalar ratio $r=0.001$. Two CIB flux cuts ($S_1$ and $S_2$) are applied to the CIB-related power spectrum calculations. (Right) CIB polarization power spectra at 857 GHz. At this frequency, the CIB is the dominating fluctuation and its $B$-mode signal becomes comparable to that of the CMB lensing. Future space submillimetre experiments could detect the CIB $E$- and $B$-mode signals.}\label{pol4}
\end{figure*}

\begin{figure*}
\includegraphics[width=19cm, height=9.5cm]{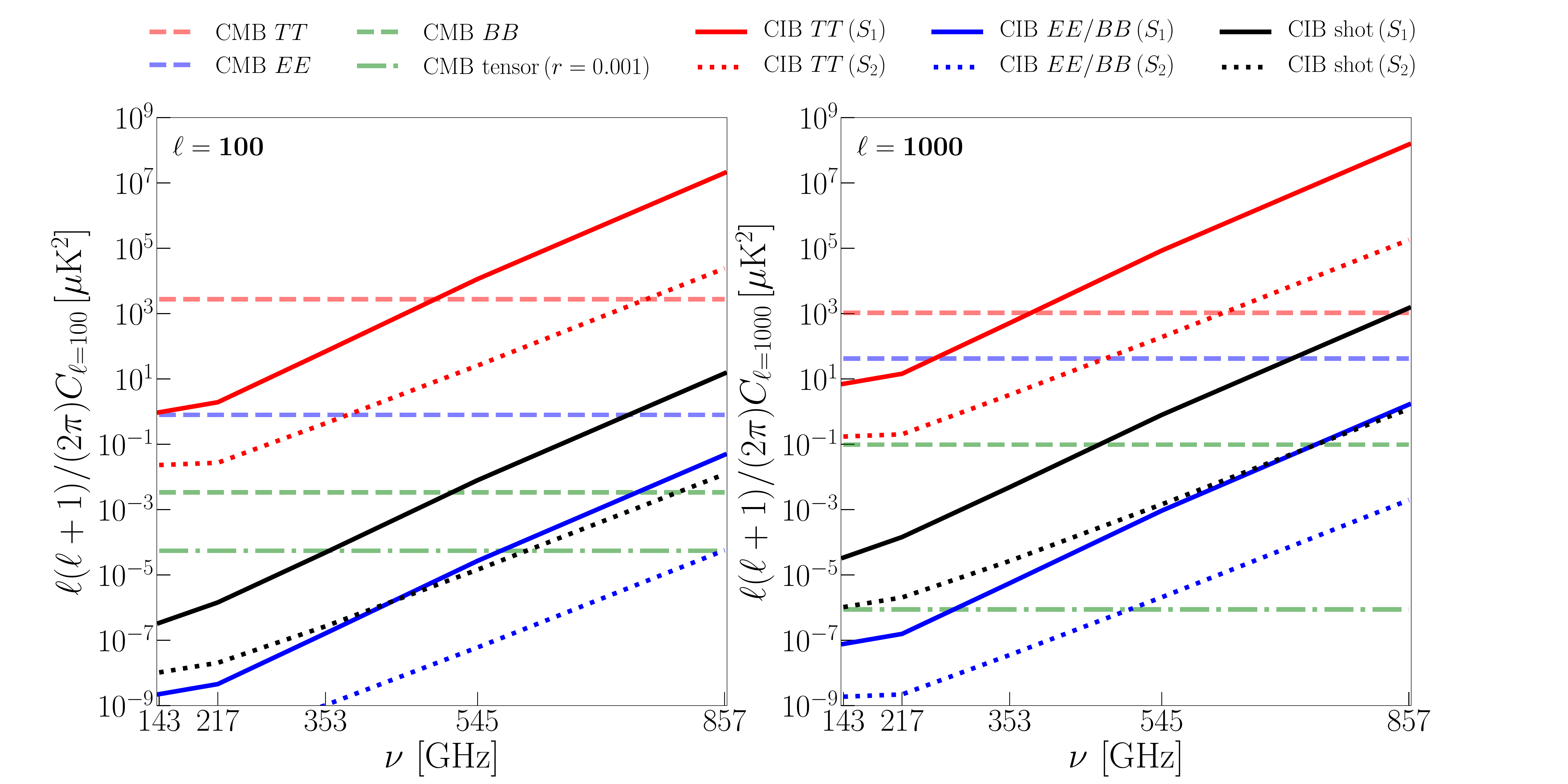}
\caption{CIB polarization power with respect to frequencies. Two cases are evaluated at $\ell=100$ (left) and $\ell$=1000 (right) at which the inflationary $B$ modes and gravitational lensing $B$ modes have peak contributions. It is seen that the CIB $B$-mode signal will exceed the CMB lensing $B$-mode at higher frequencies $\nu>545$ GHz, but will not contaminate the CMB inflationary $B$-mode signal at $\nu<353$ GHz.}\label{pp}
\end{figure*}

\begin{eqnarray}
P^{EE/BB}(k,z)&=&F^2(z)\langle p^2\rangle\int \frac{d^3k_1}{(2\pi)^3}[f_{A/B}(\bk_1)+f_{A/B}(\bk_2)]\nonumber\\
&&\times f_{A/B}(\bk_2)P^m(k_1)P^m(k_2),\label{3dpol}
\end{eqnarray}
where $P^m(k)$ is the 3D matter power spectrum. The fiducial parameter set is $\{A, C_1, \Omega_m, H_0\}$ = $\{1, 5\times 10^{-14}h^{-2} \rm{M}^{-1}_{\odot}\rm{Mpc}^{3},  0.27, 67.5\,{\rm km/s/Mpc}\}$ with a reduced Hubble constant $h=H_0/100$. The $k$ cutoff in Eq. (\ref{3dpol}) is set to 2$\rm {Mpc}^{-1}$ which is sufficiently large for convergence. This 3D polarization power spectrum can be also computed using a halo model \citep{2010MNRAS.402.2127S}.

To calculate the CIB intensity and polarization power spectra at a broad range of frequencies, we make a few approximations. The CIB source redshift distributions $dN/dz$ are poorly constrained at lower CMB frequencies, especially 217 and 143 GHz. Instead of using various empirical redshift distributions at different frequencies, we adopt the same shape of $dN/dz$ for a lower frequency but allow the amplitude to change. Specially, we derive a ratio of the Planck CIB temperature power spectrum at frequency $\nu$ to the one at 857 GHz at $\ell=900$ \citep{2014A&A...571A..30P} and scale the $dN/dz$ at 857 GHz by this factor. As a test of this approximation, we perform calculations with an empirical 545 redshift distribution and find that a good agreement is reached.

Due to the discrete nature of the CIB sources and limited flux sensitivity, the shot noise should be taken into account and be determined by a flux threshold $S_{\rm max}$ of an experiment. In this work, we adopt two different flux cuts for the power spectrum calculation, corresponding to 5\% ($S_1$) and 50\% ($S_2$) source masking, as derived from source counts $\int_0^{S_{\rm max}} dS dN/dS$ at each frequency. Specifically, they are $S_1$ = $\{0.14, 0.30, 0.88, 2.20, 4.01\}$ {\rm mJy} and $S_2$ = $\{0.02, 0.03, 0.05, 0.08, 0.09\}$ {\rm mJy} at frequencies 143, 217, 353, 545 and 857 GHz. We note that the flux distribution $dN/dS$ with frequency is model dependent \citep{2003MNRAS.338..555L, 2009MNRAS.394..117R, 2010MNRAS.409..109G} and the flux cuts may vary with models. We use the flux distributions $dN/dS$ from \cite{2011A&A...529A...4B}, and calculate the shot noise at each frequency $\nu$ with the flux cut $S_{\rm max}$ as
\begin{equation}
C^{({\rm shot})}_{\ell}=\langle p^2\rangle\int_0^{S_{\rm max}}dS S^2\frac{dN}{dS}.
\end{equation}

Two different shot-noise levels are compared at the flux cuts $S_1$ and $S_2$. The average galaxy polarization fraction $\langle p\rangle=0.01$ is assumed. Similarly, the mean intensities of CIB fluctuations are adjusted according to the flux cuts so both the CIB temperature and polarization power spectra are shifted by the flux cuts as well. In Fig. \ref{pol4} (left), we show the theoretical power spectra of the CIB polarization at $\langle p\rangle=0.01$ from frequencies 143 GHz to 545 GHz. One remarkable feature of the CIB polarization is that the IA produces equal power on the $E$ and $B$ modes, whereas for the Galactic dust, the ratio of $E$ to $B$ is $\sim 2$ \citep{2018arXiv180104945P}. The shape of the polarization power spectrum is non-white at small scales. The power of CIB polarization anisotropy $\sqrt{\ell(\ell+1)/(2\pi) C_{\ell}}$ is about four orders of magnitude fainter than its intensity anisotropy, whereas the CMB polarization is relatively much brighter --- one order of magnitude fainter than its temperature. The calculation of the $B$-mode power spectrum indicates that the inflationary CMB $B$-model signal will not be contaminated at observing frequencies $\nu<353$ GHz and if the tensor-to-scalar ratio $r>0.001$.

We show the polarization power spectra at 857 GHz, where CIB emission is expected to dominate the extragalactic sky, in Fig. \ref{pol4} (right). As seen from Figs. \ref{pol4} and \ref{pp}, the CIB $B$-mode signal is increasing as the frequency increases and becomes comparable to the CMB lensing $B$-mode signal at 857 GHz. The measured signal is a sum of the IA-induced polarization signal, shot noise and instrumental noise. Polarization signals of other secondaries, such as the polarized tSZ, are ignored in this work. Also, Galactic foregrounds will not be a problem for the CIB polarization detection. Optical surveys in the future will make direct measurements of the intrinsic alignment which can be cross-correlated with the CIB polarization data to more robustly detect the CIB polarization signals. 

\section{Conclusions}
\label{con}

In this work, we establish a theoretical model for polarization of the cosmic infrared background, in which polarization directions of individual galaxies are aligned with tidal fields. Theoretical calculations show that both the CIB $E$ and $B$ modes are created with an equal power that is about four orders of magnitude less than the CIB intensity anisotropy. The CIB $B$-model signal will not become a concerning foreground for the CMB inflationary $B$-mode searches at frequencies $<353$ GHz. However, at the CIB dominated frequencies, such as 545 and 857 GHz, the CIB polarization signals could be detected by space submillimetre observations in future, and could become a new probe of structure formation.

\section{Acknowledgments}
This research was supported in part by Perimeter Institute for Theoretical Physics. Research at Perimeter Institute is supported by the Government of Canada through the Department of Innovation, Science, and Economic Development, and by the Province of Ontario through the Ministry of Research and Innovation. This research is supported by the Brand and Monica Fortner Chair.\\

\bibliography{cibpol}

\end{document}